\documentclass[a4paper]{article}

\usepackage[margin=1in]{geometry}

\usepackage[T1]{fontenc}
\newcommand{\changefont}[3]{
\fontfamily{#1} \fontseries{#2} \fontshape{#3} \selectfont}

\changefont{ptm}{m}{n}

\usepackage{setspace} 

\usepackage{graphicx,epstopdf}

\usepackage{amsfonts,amssymb,amsmath,amsgen,amsopn,amsbsy,theorem,graphicx,epsfig}
\usepackage{graphics}
\usepackage{mathrsfs}
\usepackage{color}

\newtheorem{theorem}{Theorem}[section]

\long\def\symbolfootnote[#1]#2{\begingroup%
\def\thefootnote{\fnsymbol{footnote}}\footnote[#1]{#2}\endgroup} 

\begin{document}

\begin{center}
\Large \textbf{Persistence and Doubling of Chaotic Attractors in Coupled 3-Cell Hopfield Neural Networks}
\end{center}

\begin{center}
\normalsize \textbf{Mehmet Onur Fen$^{a,}$\symbolfootnote[1]{Corresponding Author. Tel: +90 312 585 0217, E-mail: monur.fen@gmail.com, onur.fen@tedu.edu.tr}, Fatma Tokmak Fen$^{b}$} \\
\vspace{0.2cm}
\textit{\textbf{\footnotesize$^a$Department of Mathematics, TED University, 06420 Ankara, Turkey}} \\
\textit{\textbf{\footnotesize$^b$Department of Mathematics, Gazi University, 06560 Ankara, Turkey}} 
\vspace{0.1cm}
\end{center}

\vspace{0.3cm}

\begin{center}
\textbf{Abstract}
\end{center}

\noindent\ignorespaces
Two novel phenomena for unidirectionally coupled $3$-cell Hopfield neural networks (HNNs) are investigated. The first one is the persistence of chaos, which means the permanency of sensitivity and infinitely many unstable periodic oscillations in the response HNN even if the networks are not synchronized in the generalized sense. Doubling of chaotic attractors is the second phenomenon realized in this study. It can be achieved when the response network possesses two stable point attractors in the absence of the driving. This feature leads to the formation of two coexisting chaotic attractors with disjoint basins. Lyapunov functions are utilized to deduce the presence of an invariant region, and the sensitivity is rigorously proved. The absence of synchronization is approved via the auxiliary system approach and analysis of conditional Lyapunov exponents. Additionally, quadruple and octuple coexisting chaotic attractors are demonstrated, and the formation of hyperchaos is discussed. 

\vspace{0.2cm}
 
\noindent\ignorespaces \textbf{Keywords:} Hopfield neural networks; Persistence of chaos; Doubling of chaotic attractors; Sensitivity; Generalized synchronization; Lyapunov functions 

\vspace{0.6cm}

\section{Introduction} \label{sec1}

Impressed by the sensory processing of the brain, which comprises a highly interconnected network of neurons, researches for artificial neural networks mainly focus on revealing and imitating its functioning \cite{Krogh08,Wang07}. A class of deterministic artificial neural network model was proposed by Hopfield \cite{Hopfield84} in 1984. From the mathematical point of view, dynamics of Hopfield neural networks (HNNs) are governed by systems of nonlinear differential equations. HNNs have applications in various fields of science and engineering such as optimization, cryptography, image segmentation, medical imaging, and image restoration \cite{Joya02}-\cite{Abbiss94}. Moreover, they are capable of producing several dynamical phenomena such as regular oscillations including periodic, almost periodic, and anti-periodic ones, bursting oscillations, period-doubling bifurcations, antimonotonicity, chaos, and hyperchaos \cite{Liu07}-\cite{Xu18}. 

Based on the analyses of electroencephalography signals, the studies \cite{Freeman87}-\cite{Yambe05} reveal that chaos is an indispensable part of neural activities in the brain. Chaos can be observed not only in healthy brain functioning but also in the case of diseases. For instance, it was mentioned in \cite{Babloyantz86} that the time series obtained from a human during epileptic seizure has a positive Lyapunov exponent and displays chaotic behavior. For that reason, chaotic artificial neural networks can be useful to perform brain activity simulations in a more realistic way, and researches in this context are crucial. 

Motivated by the potential role of chaos in neuroscience and utility of artificial neural networks for imitating brain dynamics, in the present paper we study the concepts of persistence and doubling of chaotic attractors in HNNs. To the best of our knowledge this is the  first time in the literature that these phenomena are realized for neural networks. The model under investigation consists of a couple of $3$-cell networks, called the drive and response, with a skew product structure in which the drive is chaotic. Persistence of chaos means the permanency of chaotic behavior in the response HNN, even if the networks are not synchronized in the generalized sense. In this phenomenon a unidirectional coupling is established between two chaotic HNNs, and it is theoretically approved that the response proceeds to display chaos. On the other hand, we achieve doubling of chaotic attractors when the response network possesses two stable equilibrium points in the absence of the driving. It is revealed that two chaotic attractors, which have disjoint basins of attraction, coexist in the resultant dynamics. In other words, the chaotic attractor of the drive network is doubled. This procedure can be applied to chains of unidirectionally coupled HNNs, and in this context the coexistence of four and eight chaotic attractors are also demonstrated. 

In the literature chaos in coupled systems has been studied within the scope of synchronization \cite{Miranda04}. This phenomenon was revealed by Pecora and Carroll \cite{Pecora90} for identical systems, and its generalization for non-identical ones was proposed by Rulkov et al. \cite{Rulkov95}. Generalized synchronization (GS) is said to occur in unidirectionally coupled systems if there is a transformation or synchronization function between their states when the transients are died away \cite{Rulkov95}-\cite{Hunt97}. A necessary and sufficient condition for the presence of GS in terms of asymptotic stability was provided by Kocarev and Parlitz \cite{Kocarev96}. Additionally, results on the continuity and smoothness of the synchronization function were attained in \cite{Hunt97,Afraimovich01}. Although synchronization is a regime in which the response proceeds to exhibit chaotic behavior \cite{Abarbanel96}, this is questionable in its absence. In this study, the concept of persistence of chaos is asserted to give an answer to this question. It is shown that the response network maintains to produce chaos regardless of GS. 

Lyapunov functionals \cite{He92}, mutual false nearest neighbors \cite{Rulkov95}, auxiliary system approach \cite{Abarbanel96}, and analysis of conditional Lyapunov exponents \cite{Pecora90, Kocarev96} are some of the techniques practical to detect GS. If there is at least one nonnegative conditional Lyapunov exponent, then one can make mention of its absence \cite{Miranda04}. In the auxiliary system approach, on the other hand, one considers an identical copy of the response, called the auxiliary system, and determines the stability of the manifold of identical oscillations in the combined phase space of the response and auxiliary systems by taking into account different initial data from the basin of attraction. The stability of this manifold implies the stability of the synchronization manifold, and vice versa \cite{Abarbanel96}. Thus, the instability of the diagonal manifold ensures the absence of GS. The observation of a stable regime of identical outputs of the response and auxiliary systems could fail provided there are multiple basins of attraction \cite{Abarbanel96}. In the present study both conditional Lyapunov exponents and auxiliary system approach are utilized to confirm the absence of GS.

To obtain persistence and doubling of chaotic attractors, weak coupling strengths between the drive and response HNNs are utilized.  Theorem \ref{unifboundedness} given in Section \ref{Sec2} guarantees the existence of an invariant region for the response network provided that the coupling is weak. Additionally, in the case of doubling of chaotic attractors, weak couplings ensure the basins of attraction to be disjoint. This is in concordance with the permanency of stable manifolds of flows under small perturbations \cite{Hirsch1977}. It was mentioned in \cite{Izhikevich99} that the involvement of a rhythmic activity of few cells with low firing rates may cause weak connections in neural ensembles. Such activities exist, for instance, in the rat hippocampus \cite{Buzsaki92}. On the other hand, chaotic dynamics have been observed in that part of the rat brain \cite{Koch92,Slutzky01}, even though weak connections may take place. Moreover, due to aging, chemicals, and diseases disruptions in synaptic connections can occur \cite{Wang07, Tsai04, Tracht97, Teter02}, and this may result in weakening coupling strengths. In this regard, the investigation of chaotic dynamics in weakly coupled networks is an interesting task.

The study \cite{Fen14} was concerned with the entrainment of limit cycles and tori by chaos. This concept was achieved by coupling a chaotic HNN with another one possessing an attracting limit cycle or torus in the absence of driving. The main difference of the present paper compared to \cite{Fen14} is in the structure of the response network. Before the coupling is established, we require the response HNN to possess either a chaotic attractor or two stable equilibrium points. It is shown that these types of dynamical behavior respectively lead to the occurrence of the persistence and doubling of chaotic attractors. On the other hand, the coexistence of multiple chaotic attractors in HNNs was demonstrated also in the papers \cite{Nijitacke18,Bao17,Zheng10b,Rajagopal18}. In these studies, coupled networks were not taken into account and chaos was demonstrated as an intrinsic property. However, we achieve doubling of chaotic attractors by coupling two HNNs, and the chaotic behavior displayed by the response is an extrinsic feature since the source of chaos is the drive. 

The main contributions of the present study are as follows.
\begin{itemize}
\item[i.] The permanency of chaotic behavior in the response network is achieved when a unidirectional coupling is established between two chaotic $3$-cell HNNs. One of the novelties is the realization of this phenomenon even if GS does not take place.
\item[ii.] It is shown that the proposed coupling scheme can lead to the formation of a hyperchaotic attractor in a $6$-cell HNN such that there are two positive Lyapunov exponents.  
\item[iii.] This is the first time in the literature that the doubling of chaotic attractors in HNNs is demonstrated via coupling. Coexisting double, quadruple and octuple chaotic attractors are respectively achieved for $6$-cell, $9$-cell and $12$-cell HNNs.
\item[iv.] The proposed coupling scheme provides an opportunity to benefit from Lyapunov functions for approving the existence of a positively invariant region as well as uniform ultimate boundedness. The latter allows to deduce the presence of infinite number of periodic outputs, which are unstable.
\item[v.] The presence of sensitivity, which means the divergence of initially nearby outputs, is confirmed rigorously when the response HNN admits a chaotic attractor or two stable equilibrium points in the absence of driving. This feature is also simulated numerically.
\end{itemize}

The rest of the paper is organized as follows. Section \ref{Sec2} is devoted to the description of our model and related preliminary concepts such as the presence of a positively invariant region and sensitivity. The verification of boundedness is an important task from the applications point of view \cite{Peng05}, and to guarantee this feature we perform a theoretical discussion based on Lyapunov functions. Section \ref{Sec3}, on the other hand, is concerned with persistence of chaos in HNNs. It is demonstrated that even in the absence of GS the chaotic behavior of the response network is permanent. The doubling of chaotic attractors, which leads to the coexistence of multiple chaotic attractors, is discussed in Section \ref{Sec4}. This is performed by setting up a coupling between a chaotic HNN and another one possessing two stable equilibrium points. Multiple chaotic attractors with disjoint basins are illustrated. Finally, some concluding remarks are mentioned in Section \ref{Sec5}, and a rigorous proof for sensitivity in the response dynamics is provided in the Appendix.

\section{The model and theoretical foundations} \label{Sec2}

An HNN with $n$ neurons can be described by the system of differential equations 
\begin{eqnarray} \label{HNNmain}
C_i \displaystyle \dot{x}_i  = - \frac{x_i}{R_i} + \displaystyle \sum_{j=1}^{n} w_{ij} \tanh(x_j) + I_i, \ i=1,2,\ldots,n,
\end{eqnarray}
where the constants $C_i>0$ and $R_i>0$ are respectively the capacitance and resistance between the outside and inside of the cell membrane of the $i$th neuron, $x_i$ is the membrane voltage of the $i$th neuron, $I_i$ is the input bias current, the activation function $\tanh(x_j)$ represents the voltage input from the $j$th neuron, and $w_{ij}$ is the synaptic connection weight between the $i$th and $j$th neurons \cite{Hopfield84,Peng05,Cao2001,Alonso09,Mathias}. In this paper we take into account $3$-cell HNNs, that is, we study the case $n=3$. One can express network (\ref{HNNmain}), in this instance, in the form
\begin{eqnarray} \label{drive}
\dot{x} =-Cx+Wf(x)+I,
\end{eqnarray}
where $x=(x_1,x_2, x_3)^T$ is the neuron state vector, $C=\textrm{diag}(c_1, c_2, c_3)$ is a diagonal matrix with $c_i=\displaystyle \frac{1}{C_iR_i}$, $i=1,2,3$, $$f(x)=(\tanh(x_1), \tanh(x_2), \tanh(x_3))^T,$$  $W=(w_{ij}/C_i)_{3 \times 3}$ is the connection matrix,  and $I=(I_1/C_1, I_2/C_2, I_3/C_3)^T$ is the external input vector. 

Our main subject of investigation is coupled HNNs consisting of a drive network of the form (\ref{drive})
and a $3$-cell response network whose dynamics is governed by 
\begin{eqnarray} \label{response}
\dot{y}=-\widetilde{C}y+\widetilde{W}f(y) + \lambda g(x(t)), 
\end{eqnarray}
where $y=(y_1,y_2, y_3)^T$ is the neuron state vector, $\widetilde{C}=\textrm{diag}(\widetilde c_1, \widetilde c_2, \widetilde c_3)$ with $\widetilde c_i>0$ for each $i=1,2,3$, $\widetilde{W}$ is a $3 \times 3$ connection matrix, $\lambda$ is a nonzero real constant which determines the coupling strength, $x(t)=(x_1(t),x_2(t), x_3(t))^T$ is an output of (\ref{drive}), $$f(y)=(\tanh(y_1), \tanh(y_2), \tanh(y_3))^T,$$ and 
$$g(x) = (g_1(x), g_2(x), g_3(x))^T,$$
in which $g_1$, $g_2$, and $g_3$ are continuous real-valued functions. 
It is worth noting that the response HNN (\ref{response}) is obtained by implementing the term $\lambda g(x(t))$ as an external input to the HNN
\begin{eqnarray} \label{absencecase}
\dot{u}=-\widetilde{C}u+\widetilde{W}f(u). 
\end{eqnarray}

We mainly assume that the drive HNN (\ref{drive}) possesses a chaotic attractor. Under this assumption, there is a compact set $\Lambda_x \subset \mathbb R^3$ which is invariant under the flow of (\ref{drive}). In the rest of the paper we make use of the Euclidean norm for vectors and the spectral norm for square matrices.

In what follows we will refer to solutions of (\ref{drive}) with initial data from the chaotic attractor as chaotic outputs. More precisely, an output $x(t)=(x_1(t),x_2(t), x_3(t))^T$ of (\ref{drive}) is said to be chaotic if $x(0)$ belongs to the chaotic attractor. The drive HNN (\ref{drive}) is called sensitive if there exist numbers $\epsilon>0$ and $\Delta>0$ such that for an arbitrary number $\delta>0$ and for each chaotic output $x(t)$ of (\ref{drive}), there exist a chaotic output $\widetilde{x}(t)$ of the same network and an interval $J \subset [0,\infty)$ with a length no less than $\Delta$ satisfying $\left\|x(0)-\widetilde{x}(0)\right\|<\delta$ and $\left\|x(t)-\widetilde{x}(t)\right\| > \epsilon$ for every $t \in J$ \cite{Fen14,Akhmet2015}.

Our another assumption is the existence of a bounded subset $\Lambda_y$ of $\mathbb R^3$ which is invariant under the flow of the response HNN (\ref{response}) for every chaotic output $x(t)=(x_1(t),x_2(t), x_3(t))^T$ of the drive (\ref{drive}). For a given output $x(t)$ of  (\ref{drive}), let us denote by $\varphi_{x}(t,t_0,\alpha)$ the output of (\ref{response}) satisfying the initial data $\varphi_{x}(t_0,t_0,\alpha)=\alpha$, where $t_0\geq 0$ and $\alpha \in \mathbb R^3$. 

The outputs of the response HNN (\ref{response}) are uniformly ultimately bounded if there exists a number $B_0>0$ and corresponding to any number $\gamma > 0$ there exists a number $T(\gamma)>0$ such that $\left\|\alpha\right\| \leq \gamma$ implies that for each output $x(t)$ of the drive network (\ref{drive}) and $t_0 \geq 0$, we have $\left\|\varphi_x(t,t_0,\alpha)\right\|<B_0$ for every $t \geq t_0+T(\gamma)$ \cite{Yoshizawa66}.

The following result is a restatement of Theorem 1 provided in paper \cite{Akhmet2015}, and it can be utilized to show the existence of an invariant set for the response network (\ref{response}). 

\begin{theorem}\label{unifboundedness} (\cite{Akhmet2015})
Suppose that the function $g(x)$ is bounded and $V(u)$ is a Lyapunov function defined on $\mathbb R^3$ which has continuous first order partial derivatives such that for some $B \geq 0$ the following conditions are satisfied on the region $\left\|u\right\| \geq B:$
\begin{itemize}
\item[i.] $V(u) \geq a(\left\|u\right\|)$, where  $a(r)$ is a continuous, increasing function defined for $r \geq B$ which satisfies $a(B)>0$ and $a(r) \to \infty$ as $r\to\infty$;
\item[ii.] $\dot{V}_{(\ref{absencecase})}(u) \leq - b\left(\left\|u\right\|\right)$, where $b(r)$ is an increasing function defined for $r\geq B$ which satisfies $b(B)>0$; 
\item[iii.] $\displaystyle \Big\| \frac{\partial V}{\partial u}(u)\Big\| \leq c\left(\left\|u\right\|\right)$, where $c(r)$ is a function defined for $r \geq B$, and there exists a number $\mu>0$ such that $0 < c(r) \leq \mu b(r)$ for all $r \geq B$.
\end{itemize}
Then, for sufficiently small $\left|\lambda\right|$, the outputs of the response network (\ref{response}) are uniformly ultimately bounded.
\end{theorem}

If the conditions of Theorem \ref{unifboundedness} are fulfilled, then Theorem 15.8 provided in \cite{Yoshizawa66} ensures that for each unstable periodic output of the drive network (\ref{drive}), the response (\ref{response}) produces a periodic output with the same period. In this case, the instability in (\ref{drive}) implies the instability of the corresponding periodic output of the coupled network (\ref{drive})-(\ref{response}). Accordingly, one can confirm that if the drive (\ref{drive}) possesses infinitely many unstable periodic outputs, then the same is true for the coupled network (\ref{drive})-(\ref{response}). This kind of behavior can be regarded as one of the ingredients of chaos \cite{Wiggins88}.

The response HNN (\ref{response}) is said to admit sensitivity if there exist numbers $\widetilde{\epsilon} >0$ and $\widetilde{\Delta}>0$ such that for an arbitrary number $\widetilde{\delta}>0$, each $\alpha_1 \in \Lambda_y$, and each chaotic output $x(t)$ of (\ref{drive}), there exist $\alpha_2 \in \Lambda_y$, a chaotic output $\widetilde{x}(t)$ of (\ref{drive}), and an interval $\widetilde{J} \subset [0,\infty)$ with a length no less than $\widetilde{\Delta}$ satisfying $\left\|\alpha_1-\alpha_2\right\|<\widetilde{\delta}$ and $\left\|\varphi_{x}(t,0,\alpha_1)-\varphi_{\widetilde{x}}(t,0,\alpha_2)\right\| > \widetilde{\epsilon}$ for every $t \in \widetilde{J}$ \cite{Fen14,Akhmet2015}. 

The following assumption is required.
\begin{itemize}
\item[\textbf{(A)}] There is a positive real constant $L$ such that 
$\left\|g(x) - g(\overline{x})\right\| \geq L \left\|x-\overline{x}\right\|$
for every $x, \overline{x} \in \Lambda_x$.
\end{itemize}

The presence of the sensitivity feature for (\ref{response}) is mentioned in the next theorem.

\begin{theorem} \label{thm_sensitive}
Suppose that the assumption (A) is fulfilled. Then, the response HNN (\ref{response}) admits sensitivity.
\end{theorem}
 
For the convenience of the reader, the proof of Theorem \ref{thm_sensitive} is given in the Appendix. It is remarkable that  the result of Theorem \ref{thm_sensitive} holds regardless of the dynamical state of (\ref{absencecase}). For instance, HNN (\ref{absencecase}) can possess a chaotic attractor or stable equilibrium points.  

Now, we will verify that the assumption (A) can be attained when the tangent hyperbolic function is used for establishing the coupling term $g(x)$ in (\ref{response}). Let us take 
\begin{eqnarray} \label{functiong}
g(x)=P(\tanh(x_1), \tanh(x_2), \tanh(x_3))^T
\end{eqnarray} 
for some $3 \times 3$ nonsingular matrix $P$, where $x=(x_1,x_2,x_3)^T$. In this case, there exists a bounded region $\Omega \subset \mathbb R^3$ such that for every $x,\overline x \in \Lambda_x$ with $x=(x_1,x_2,x_3)^T$ and $\overline x =(\overline x_1,\overline x_2,\overline x_3)^T$ we have 
$$\left\|g(x)-g\left(\overline{x}\right)\right\| \geq \frac{1}{\left\|P^{-1}\right\|} \Big(\sum_{j=1}^3 \textrm{sech}^4(p_j)(x_j-\overline x_j)^2\Big)^{1/2} \geq L \left\|x-\overline x\right\|$$ for some $(p_1,p_2,p_3) \in \Omega$, where $L$ is the positive number defined by $$L=\displaystyle \frac{1}{\left\|P^{-1}\right\|}\min_{j=1,2,3}\big\{\inf_{(p_1,p_2,p_3) \in \Omega} \textrm{sech}^2(p_j) \big\}.$$
Therefore, in conformity with Theorem \ref{thm_sensitive}, the response network (\ref{response}) admits sensitivity provided that (\ref{functiong}) holds. 

We would like to point out that equation (\ref{functiong}) gives rise to the formation of a $6$-cell HNN when the coupling between the networks (\ref{drive}) and (\ref{response}) is established, and such couplings are taken into account in Sections \ref{Sec3} and \ref{Sec4}.

The next section is concerned with the persistence of chaos in coupled chaotic HNNs. When this phenomenon occurs the dynamics of the response network remains to possess chaotic outputs even in the absence of GS.

\section{Persistence of chaos}\label{Sec3} 

It was shown by Zheng et al. \cite{Zheng10} that the HNN
\begin{eqnarray} \label{persistence1}
&& \dot{x}_1=-x_1-0.6\tanh (x_1) +1.2 \tanh (x_2) -7\tanh (x_3)\nonumber \\
&& \dot{x}_2=-x_2 + 1.1 \tanh (x_1) -0.1 \tanh (x_2) + 2.8 \tanh (x_3) \nonumber  \\
&& \dot{x}_3=-x_3 +0.8 \tanh (x_1) -2 \tanh (x_2) +4 \tanh (x_3) 
\end{eqnarray}
has a positive Lyapunov exponent and admits a double-scroll chaotic attractor. Considering (\ref{persistence1}) as the drive, we set up the response network
\begin{eqnarray} \label{persistence2}
 \dot{y}_1&=&-y_1+3.4\tanh (y_1) -1.6 \tanh (y_2) +0.7\tanh (y_3) +0.034 \tanh(x_1) \nonumber \\ && -0.008\tanh(x_2) \nonumber \\
 \dot{y}_2&=&-y_2 + 2.5 \tanh (y_1) + 0.95 \tanh (y_3) +0.012\tanh(x_1)+0.023\tanh(x_2) \nonumber  \\ && -0.016\tanh(x_3) \nonumber  \\
 \dot{y}_3&=&-y_3 -12.5 \tanh (y_1) +0.5 \tanh (y_2) +0.003\tanh(x_2) +0.042\tanh(x_3),   
\end{eqnarray}
in which $x=(x_1,x_2,x_3)^T$ is an output of (\ref{persistence1}). HNN (\ref{persistence2}) is originated from the usage of 
\begin{eqnarray*} \label{persistenceinput}
\begin{pmatrix}
0.034 \tanh(x_1)-0.008\tanh(x_2) \\
0.012\tanh(x_1)+0.023\tanh(x_2)-0.016\tanh(x_3) \\
0.003\tanh(x_2) +0.042\tanh(x_3)
\end{pmatrix} 
\end{eqnarray*}
as an input for the HNN
\begin{eqnarray} \label{persistence3}
&& \dot{u}_1=-u_1+3.4\tanh (u_1) -1.6 \tanh (u_2) +0.7\tanh (u_3)\nonumber \\
&& \dot{u}_2=-u_2 + 2.5 \tanh (u_1) + 0.95 \tanh (u_3) \nonumber \\
&& \dot{u}_3=-u_3 -12.5 \tanh (u_1) +0.5 \tanh (u_2).  
\end{eqnarray}
According to the results of the study \cite{Huang08}, HNN (\ref{persistence3}) possesses a chaotic attractor and its largest Lyapunov exponent is $0.1235$.

In order to verify the existence of an invariant region for network (\ref{persistence2}), we consider the Lyapunov function 
\begin{eqnarray} \label{lypfunc}
V(u)=\frac{1}{2}\left(u_1^2 + u_2^2 + u_3^2\right),
\end{eqnarray}
where $u=(u_1,u_2,u_3)^T$. One can confirm that
$V(u)=a\left(\left\|u\right\|\right)$, where $a(r)=r^2/2$. On the other hand, by means of the inequality $\left|\tanh(u_i)\right|<1$, $i=1,2,3$, we attain that
\begin{eqnarray*}
\dot{V}_{(\ref{persistence3})}(u) & =& u_1 \dot{u}_1+u_2 \dot{u}_2+u_3 \dot{u}_3 \\
&=& u_1\left(-u_1+3.4\tanh (u_1) -1.6 \tanh (u_2) +0.7\tanh (u_3)\right) \\
&+& u_2\left(-u_2 + 2.5 \tanh (u_1) + 0.95 \tanh (u_3)\right) \\
&+& u_3 \left(-u_3 -12.5 \tanh (u_1) +0.5 \tanh (u_2)\right) \\
&\leq& - \left\|u\right\|^2 + 5.1 \left|u_1\right| + 3.45 \left|u_2\right| + 13 \left|u_3\right| \\
&\leq &  -  \left\|u\right\|^2 + 13\sqrt 3 \left\|u\right\|.
\end{eqnarray*}
Therefore, $\dot{V}_{(\ref{persistence3})}(u) \leq - b\left(\left\|u\right\|\right)$ on the region $\left\|u\right\|\geq B$, in which $b(r)=\displaystyle \frac{1}{2}r^2$ and $B=26\sqrt 3$. Moreover, $\left\|\displaystyle \frac{\partial V}{\partial u}(u)\right\|=c\left(\left\|u\right\|\right)$ with $c(r)=r$. If $r \geq B$, then the inequality $0<c(r)\leq \mu b(r)$ holds, where $\mu=\displaystyle \frac{1}{13\sqrt 3}$. Hence, the conditions of Theorem \ref{unifboundedness} are fulfilled. This ensures
the uniformly ultimate boundedness of the outputs of HNN (\ref{persistence2}) as well as the presence of an invariant region. Moreover, (\ref{persistence2}) admits sensitivity in accordance with Theorem \ref{thm_sensitive}.

\begin{figure}[ht!] 
\centering
\includegraphics[width=11.5cm]{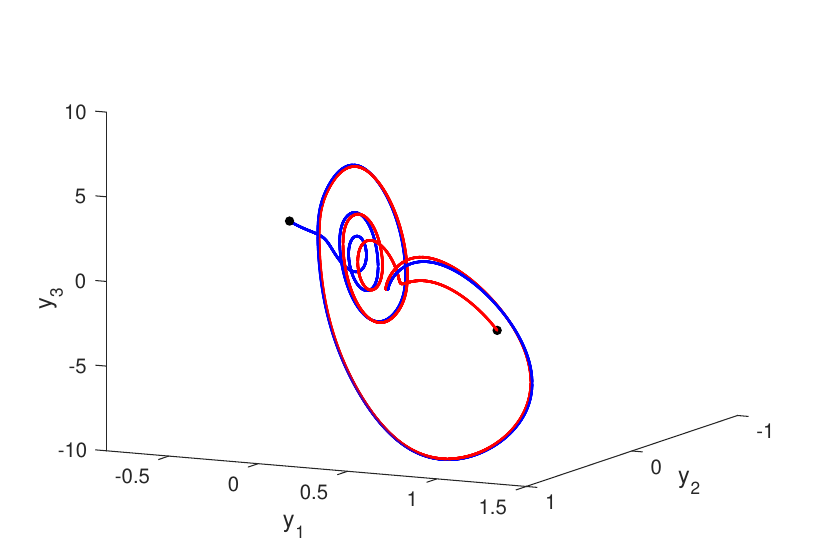}
\caption{Sensitivity in the response HNN (\ref{persistence2}). The figure shows that two initially nearby outputs of (\ref{persistence2}) eventually diverge. The thick black dots represent the positions of the outputs at $t=16.4$.}
\label{fig1n}
\end{figure}

\begin{figure}[ht!] 
\centering
\includegraphics[width=11.5cm]{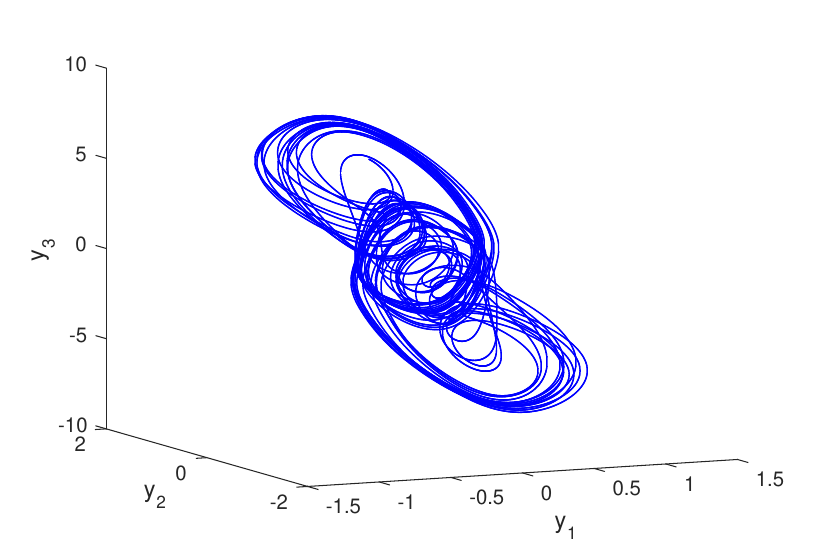}
\caption{Irregular output of the response HNN (\ref{persistence2}). The figure represents the $3$-dimensional projection of the trajectory of the coupled HNNs (\ref{persistence1})-(\ref{persistence2}) with the initial point $(3.27, -0.67, -1.19, -0.49, -0.34, 6.32)^T$ on the $y_1-y_2-y_3$ space. The simulation affirms the persistence of chaos in (\ref{persistence2}).}
\label{fig2n}
\end{figure}

Now, let us validate the result of Theorem \ref{thm_sensitive} for (\ref{persistence2}) by illustrating the sensitivity feature. For that purpose, we represent in Figure \ref{fig1n} the $3$-dimensional projections of the trajectories of the coupled network (\ref{persistence1})-(\ref{persistence2}) on the $y_1-y_2-y_3$ space with initial points $$(0.368, -0.713, -0.579, 0.123, -0.021, -1.741)^T$$ and $$(0.361, -0.719, -0.572, 0.114, -0.015, -1.745)^T$$ in blue and red, respectively. Using the time interval $[0, 16.4]$, it is demonstrated in Figure \ref{fig1n} that even though the trajectories are initially nearby, they eventually diverge, i.e., HNN (\ref{persistence2}) admits sensitivity. Figure \ref{fig2n}, on the other hand, depicts the irregular output of (\ref{persistence2}) corresponding to the initial data $x_1(0)=3.27$, $x_2(0)=-0.67$, $x_3(0)=-1.19$, $y_1(0)=-0.49$, $y_2(0)=-0.34$, $y_3(0)=6.32$, and it reveals the persistence of chaos in the resulting dynamics. In other words, the response network (\ref{persistence2}) proceeds to display chaotic behavior even if the unidirectional coupling is established with (\ref{persistence1}). Additionally, the $6$-cell HNN (\ref{persistence1})-(\ref{persistence2}) is hyperchaotic such that it possesses the positive Lyapunov exponents $0.14403$ and $0.09843$.

\begin{figure}[ht!] 
\centering
\includegraphics[width=9.5cm]{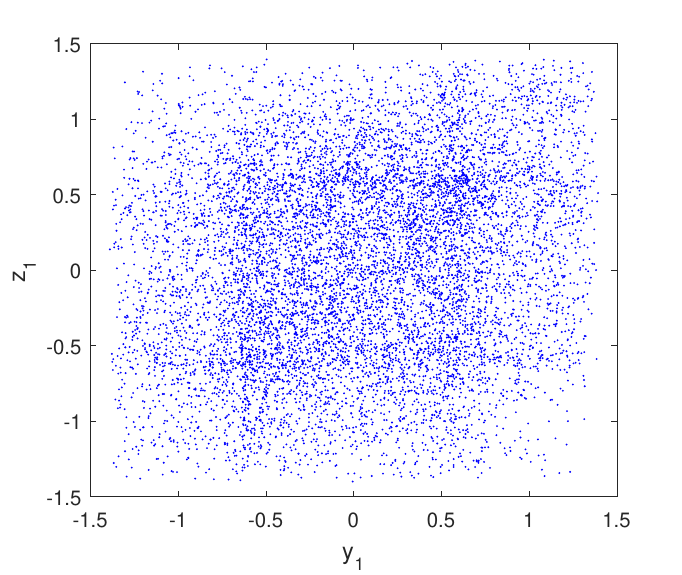}
\caption{The projection of the stroboscopic plot of (\ref{persistence1})-(\ref{persistence2})-(\ref{persistence4}) on the $y_1-z_1$ plane. The plot does not take place on the diagonal, and this manifests that the chaos is persistent even though the networks (\ref{persistence1}) and (\ref{persistence2}) are not synchronized in the generalized sense.}
\label{fig3n}
\end{figure}

The networks (\ref{persistence1}) and (\ref{persistence2}) are said to be synchronized in the generalized sense \cite{Miranda04,Rulkov95,Kocarev96} if there exist sets of initial points $\Gamma_x$, $\Gamma_y$ in $\mathbb R^3$ and a transformation $\mathcal H$, which is defined on the chaotic attractor of (\ref{persistence1}), such that the asymptotic relation
\begin{eqnarray} \label{syncheqn}
\lim_{t \to \infty} \left\|y(t)-\mathcal H (x(t))\right\| =0
\end{eqnarray}
is fulfilled for each state $x=(x_1,x_2,x_3)^T$ of (\ref{persistence1}) and $y=(y_1,y_2,y_3)^T$ of (\ref{persistence2}) starting respectively from $\Gamma_x$ and $\Gamma_y$. We will certify the absence of GS in the dynamics of the coupled network (\ref{persistence1})-(\ref{persistence2}) making use of the auxiliary system approach, which was introduced by Abarbanel et al. \cite{Abarbanel96}. In accordance with this purpose we constitute the auxiliary network
\begin{eqnarray} \label{persistence4}
 \dot{z}_1 & = & -z_1 + 3.4\tanh (z_1) -1.6 \tanh (z_2) +0.7\tanh (z_3) +0.034 \tanh(x_1) \nonumber  \\
&& -0.008\tanh(x_2) \nonumber \\
 \dot{z}_2 & = & -z_2 + 2.5 \tanh (z_1) + 0.95 \tanh (z_3) +0.012\tanh(x_1)+0.023\tanh(x_2) \nonumber \\ 
&& -0.016\tanh(x_3) \nonumber \\
 \dot{z}_3 & = & -z_3 -12.5 \tanh (z_1) +0.5 \tanh (z_2) +0.003\tanh(x_2) +0.042\tanh(x_3). 
\end{eqnarray}
It is worth noting that (\ref{persistence4}) is an identical copy of (\ref{persistence2}). Figure \ref{fig3n} shows the stroboscopic plot of the network (\ref{persistence1})-(\ref{persistence2})-(\ref{persistence4}) on the $y_1-z_1$ plane. In the simulation the initial data $x_1(0)= 3.27$, $x_2(0)=-0.67$, $x_3(0)=-1.19$, $y_1(0)=-0.49$, $y_2(0)=-0.34$, $y_3(0)= 6.32$, $z_1(0)= 0.15$, $z_2(0)= 0.27$, $z_3(0)= -2.53$ are utilized, and the first $150$ iterations are omitted to annihilate the transient motions. Because the plot in Figure \ref{fig3n} does not take place on the line $z_1=y_1$, GS does not occur. Therefore, one can confirm that chaos is persistent even though the networks (\ref{persistence1}) and (\ref{persistence2}) are unsynchronized.

Another method which is adequate to detect the presence or absence of GS is the evaluation of conditional Lyapunov exponents \cite{Pecora90,Kocarev96}. To that end, we take into account the variational system  
\begin{eqnarray} \label{persistence5}
&& \dot{\gamma}_1 = \left(-1+3.4\textrm{sech}^2 (y_1)\right) \gamma_1 -1.6 \textrm{sech}^2 (y_2) \gamma_2 +0.7\textrm{sech}^2 (y_3)\gamma_3  \nonumber \\
&& \dot{\gamma}_2 = 2.5 \textrm{sech}^2 (y_1) \gamma_1 -\gamma_2 +0.95 \textrm{sech}^2(y_3) \gamma_3 \nonumber  \\
&& \dot{\gamma}_3 = -12.5 \textrm{sech}^2 (y_1)\gamma_1 +0.5 \textrm{sech}^2 (y_2) \gamma_2 - \gamma_3.  
\end{eqnarray}
Taking advantage of the output $y=(y_1,y_2,y_3)^T$ of (\ref{persistence2}) depicted in Figure \ref{fig2n}, the largest Lyapunov exponent of (\ref{persistence5}) is evaluated as $0.089906$. The presence of a positive conditional Lyapunov exponent of (\ref{persistence2}) reveals one more time the absence of GS for the coupled HNNs (\ref{persistence1})-(\ref{persistence2}).

The next section is devoted to the coexistence of double chaotic attractors in coupled networks of the form (\ref{drive})-(\ref{response}). More precisely, we will demonstrate that if the assumption (A) is fulfilled, $|\lambda|$ is sufficiently small, and HNN (\ref{absencecase}) admits two stable equilibrium points, then the response network (\ref{response}) possesses two chaotic attractors with disjoint basins of attraction. The formations of four and eight chaotic attractors are also shown.

\section{Coexistence of multiple chaotic attractors} \label{Sec4} 

Let us take into account the network \cite{Li13}
\begin{eqnarray} \label{exdrive1}
&& \dot{x}_1=-x_1+2\tanh (x_1) - \tanh (x_2) \nonumber \\
&& \dot{x}_2=-x_2 + 1.7 \tanh (x_1) + 1.71 \tanh (x_2) + 1.1 \tanh (x_3) \nonumber \\
&& \dot{x}_3=-2x_3 -2.5 \tanh (x_1) -2.9 \tanh (x_2) + 0.56 \tanh (x_3).  
\end{eqnarray}
Making use of the topological horseshoes theory, it was verified by Li et al. \cite{Li13} that (\ref{exdrive1}) exhibits horseshoe chaos.
Next, we consider the HNN \cite{Bao17}
\begin{eqnarray} \label{exresponse1}
&& \dot{u}_1=-u_1-1.4\tanh (u_1) + 1.2 \tanh (u_2)-7\tanh(u_3) \nonumber \\
&& \dot{u}_2=-u_2 + 1.1 \tanh (u_1) + 2.8 \tanh (u_3) \nonumber \\
&& \dot{u}_3=-u_3 + k \tanh (u_1) -2 \tanh (u_2) + 4 \tanh (u_3),  
\end{eqnarray}
in which the real parameter $k$ represents the synaptic connection weight between the first and third neurons. According to the results of the study \cite{Bao17}, if $0<k \leq 0.535$, then HNN (\ref{exresponse1}) possesses a pair of coexisting equilibrium points, which are stable node-foci and symmetric about the origin.

We take $k=0.5$ in (\ref{exresponse1}), and in this case $(\pm 2.9545817, \mp  1.0128998, \mp0.9784960)$ are the stable equilibrium points of the network. By establishing unidirectional coupling between (\ref{exdrive1}) and (\ref{exresponse1}), we set up the network
\begin{eqnarray} \label{exresponse2}
 \dot{y}_1 & =& -y_1-1.4\tanh (y_1) + 1.2 \tanh (y_2)-7\tanh(y_3) + 0.032 \tanh(x_1) \nonumber \\  && + 0.012 \tanh(x_2) + 0.003 \tanh(x_3) \nonumber \\
 \dot{y}_2&=&-y_2 + 1.1 \tanh (y_1) + 2.8 \tanh (y_3) -0.008\tanh(x_1)  \nonumber \\  && +0.039\tanh(x_2)+0.005 \tanh(x_3) \nonumber \\
 \dot{y}_3&=&-y_3 + 0.5 \tanh (y_1) -2 \tanh (y_2) + 4 \tanh (y_3) + 0.006\tanh(x_1) \nonumber \\  && - 0.014 \tanh(x_2) + 0.035 \tanh(x_3),  
\end{eqnarray} 
where $x=(x_1,x_2,x_3)^T$ is an output of (\ref{exdrive1}). Here, (\ref{exdrive1}) and (\ref{exresponse2}) are respectively the drive and response networks.
 
In a similar way to Section \ref{Sec3}, using one more time the Lyapunov function (\ref{lypfunc}), it can be verified that the conditions of Theorem \ref{unifboundedness} are fulfilled with $a(r)=b(r)=\displaystyle \frac{1}{2}r^2$, $c(r)=r$, $B=19.2\sqrt 3$, and $\mu=\displaystyle \frac{1}{9.6\sqrt 3}$. This implies the uniform ultimate boundedness of the outputs of (\ref{exresponse2}) as well as the existence of an invariant region for the network due to the smallness of the coefficients of the external input terms in absolute value. According to Theorem 15.8 provided in \cite{Yoshizawa66}, for each unstable periodic output of (\ref{exdrive1}), the network (\ref{exresponse2}) generates a periodic output with the same period. Therefore, there are infinitely many unstable periodic outputs of the coupled HNNs (\ref{exdrive1})-(\ref{exresponse2}). In addition to this, Theorem \ref{thm_sensitive} guarantees the sensitivity feature of (\ref{exresponse2}).

\begin{figure}[ht!] 
\centering
\includegraphics[width=11.5cm]{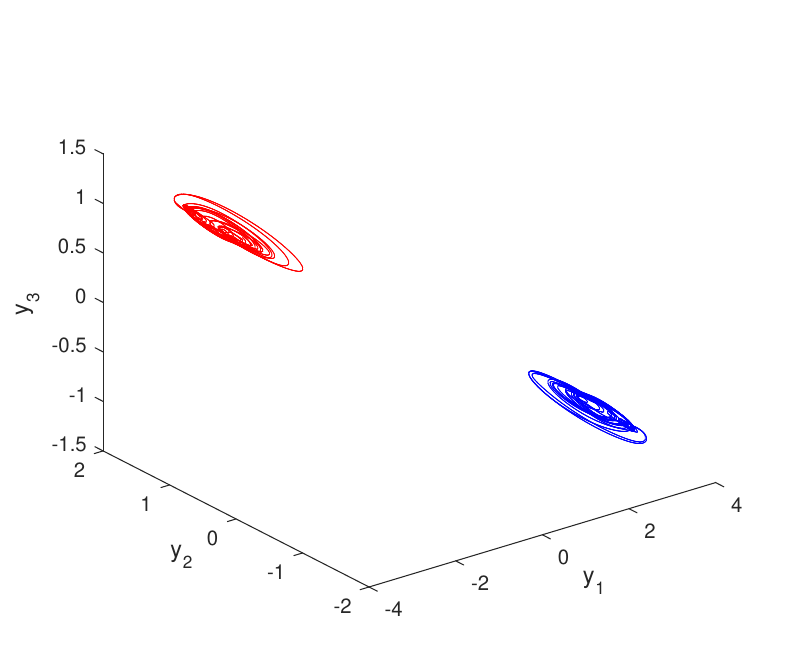}
\caption{The $3$-dimensional projections of the chaotic trajectories of the coupled HNNs (\ref{exdrive1})-(\ref{exresponse2}) on the $y_1-y_2-y_3$ space. The figure manifests the coexistence of two chaotic attractors with disjoint basins.}
\label{fig4n}
\end{figure}

\begin{figure}[ht!] 
\centering
\includegraphics[width=16.0cm]{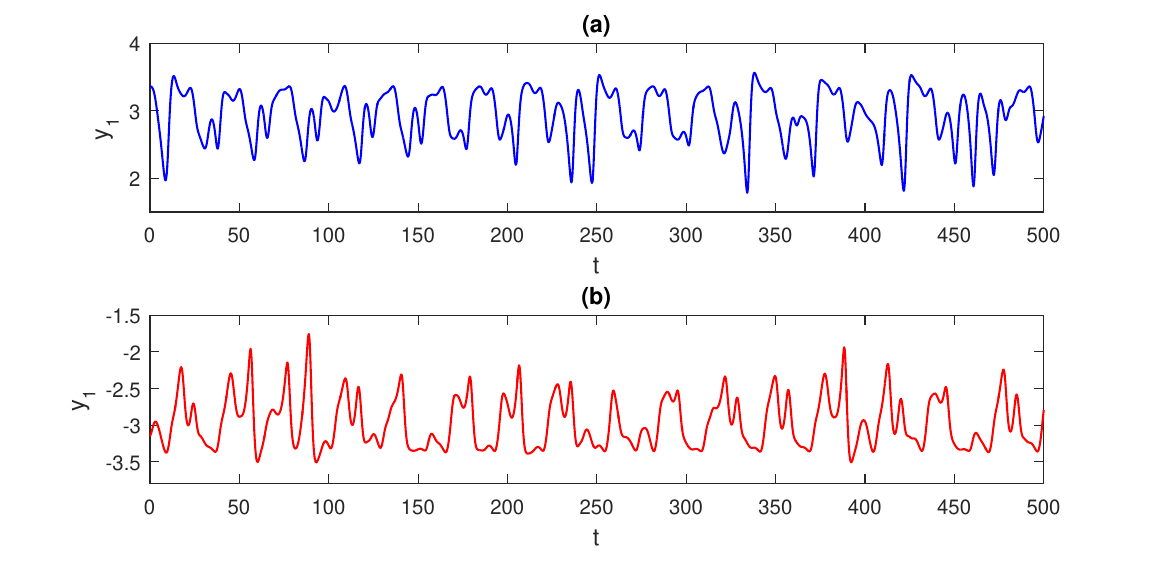}
\caption{The time-series of the $y_1$-coordinate of the response HNN (\ref{exresponse2}) corresponding to the initial data: (a)  $x_1(0)=0.15$, $x_2(0)=0.43$, $x_3(0)=-0.24$, $y_1(0)=3.36$, $y_2(0)=-1.18$, $y_3(0)=-1.16$. (b) $x_1(0)=0.07$, $x_2(0)=0.18$, $x_3(0)=0.05$, $y_1(0)=-3.15$, $y_2(0)=1.12$, $y_3(0)=1.05$. Both time-series display irregular behavior, and this certifies that two chaotic attractors take place in the dynamics.}
\label{fig5n}
\end{figure}

Figure \ref{fig4n} depicts the $3$-dimensional projections of two chaotic trajectories of the coupled HNNs (\ref{exdrive1})-(\ref{exresponse2}) on the $y_1-y_2-y_3$ space. In the figure, the trajectory in blue corresponds to the initial data $x_1(0)=0.15$, $x_2(0)=0.43$, $x_3(0)=-0.24$, $y_1(0)=3.36$, $y_2(0)=-1.18$, $y_3(0)=-1.16$, whereas the trajectory in red corresponds to $x_1(0)=0.07$, $x_2(0)=0.18$, $x_3(0)=0.05$, $y_1(0)=-3.15$, $y_2(0)=1.12$, $y_3(0)=1.05$. Besides, Figure \ref{fig5n} (a) and (b) respectively represent the time-series for the $y_1$-coordinates of the blue and red trajectories shown in Figure \ref{fig4n}. Both of the Figures \ref{fig4n} and \ref{fig5n} reveal the coexistence of two chaotic attractors in the dynamics of the coupled HNNs (\ref{exdrive1})-(\ref{exresponse2}) with disjoint basins of attraction. 

Now, in order to generate four coexisting chaotic attractors, we utilize (\ref{exresponse2}) as the drive and consider the response network
\begin{eqnarray} \label{exresponse3}
\dot{z}_1 &=& -z_1-1.4 \tanh(z_1)+1.2 \tanh(z_2)-7 \tanh(z_3)+0.086 \tanh(y_1) \nonumber \\ 
&& + 0.015 \tanh(y_3)\nonumber  \\
\dot{z}_2 &=& -z_2+1.1 \tanh(z_1)+2.8 \tanh(z_3)+0.004 \tanh(y_1)+0.121 \tanh(y_2) \nonumber \\
\dot{z}_3 &=& -z_3+0.52 \tanh(z_1)-2 \tanh(z_2)+4 \tanh(z_3) + 0.019 \tanh(y_2) \nonumber \\
&& +0.118 \tanh(y_3),  
\end{eqnarray} 
in which $y=(y_1,y_2,y_3)^T$ is an output of (\ref{exresponse2}). It is worth noting that the response network (\ref{exresponse2}) is constructed by implementing the output of (\ref{exresponse2}) as the external input for (\ref{exresponse1}) with $k=0.52$, which admits the stable equilibrium points $(\pm 2.7956906,  \mp 0.9303841,  \mp 0.9228533)$.
 
\begin{figure}[ht!] 
\centering
\includegraphics[width=14.0cm]{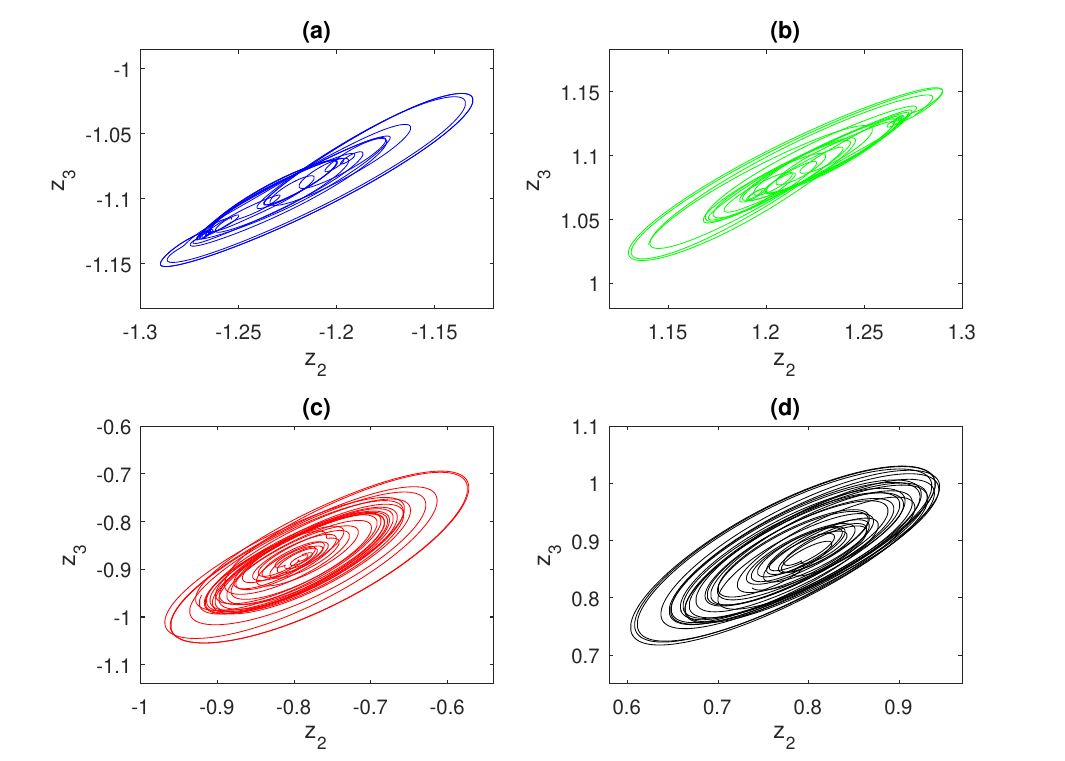}
\caption{Coexistence of four chaotic attractors in the dynamics of the $9$-cell HNN (\ref{exdrive1})-(\ref{exresponse2})-(\ref{exresponse3}). The figure depicts the $2$-dimensional projections of chaotic trajectories on the $z_2-z_3$ plane. The initial data for each of the trajectories in (a)-(d) are shown in Table \ref{table1}.}
\label{fig6n}
\end{figure}

The $2$-dimensional projections of four chaotic trajectories of the coupled HNNs (\ref{exdrive1})-(\ref{exresponse2})-(\ref{exresponse3}) on the $z_2-z_3$ plane are represented in blue, green, red, and black respectively in Figure \ref{fig6n} (a), (b), (c), and (d). The initial data corresponding to each of these trajectories are illustrated in Table \ref{table1}. Figure \ref{fig6n} shows that the dynamical behavior of (\ref{exdrive1})-(\ref{exresponse2}) is replicated around each equilibrium point of (\ref{exresponse1}), and accordingly, the $9$-cell HNN (\ref{exdrive1})-(\ref{exresponse2})-(\ref{exresponse3}) possesses four coexisting chaotic attractors. In other words, the chaotic attractors of (\ref{exdrive1})-(\ref{exresponse2}) are doubled by means of coupling with (\ref{exresponse3}). Moreover, the basins of attraction of each these chaotic attractors are disjoint. 

\begin{table}[ht]
\caption{Initial data of the trajectories represented in Figure \ref{fig6n}.}  
\centering  
\begin{tabular}{c c c c c c c c c c} 
\hline
& $x_1(0)$ & $x_2(0)$ & $x_3(0)$ & $y_1(0)$ & $y_2(0)$ & $y_3(0)$ & $z_1(0)$ & $z_2(0)$ & $z_3(0)$ \\ \hline
Figure \ref{fig6n}, (a) \ \ & $0.15$ & $0.68$ & $-1.51$ & $2.54$ & $-0.84$ & $-0.78$ & $3.37$ & $-1.27$ & $-1.13$ \\  
Figure \ref{fig6n}, (b) \ \ & $0.42$ & $0.12$ & $-0.85$ & $-2.47$ & $0.72$ & $0.97$ & $-3.13$ & $1.14$ & $1.03$ \\  
Figure \ref{fig6n}, (c) \ \ & $-0.41$ & $-0.39$ & $1.21$ & $-3.35$ & $1.19$ & $1.14$ & $2.65$ & $-0.79$ & $-0.87$ \\  
Figure \ref{fig6n}, (d) \ \ & $0.32$ & $0.38$ & $-0.92$ & $3.21$ & $-1.12$ & $-1.09$ & $-2.72$ & $0.81$ & $0.91$ \\ \hline
\end{tabular} 
\label{table1} 
\end{table} 
 
\begin{figure}[ht!] 
\centering
\includegraphics[width=13.0cm]{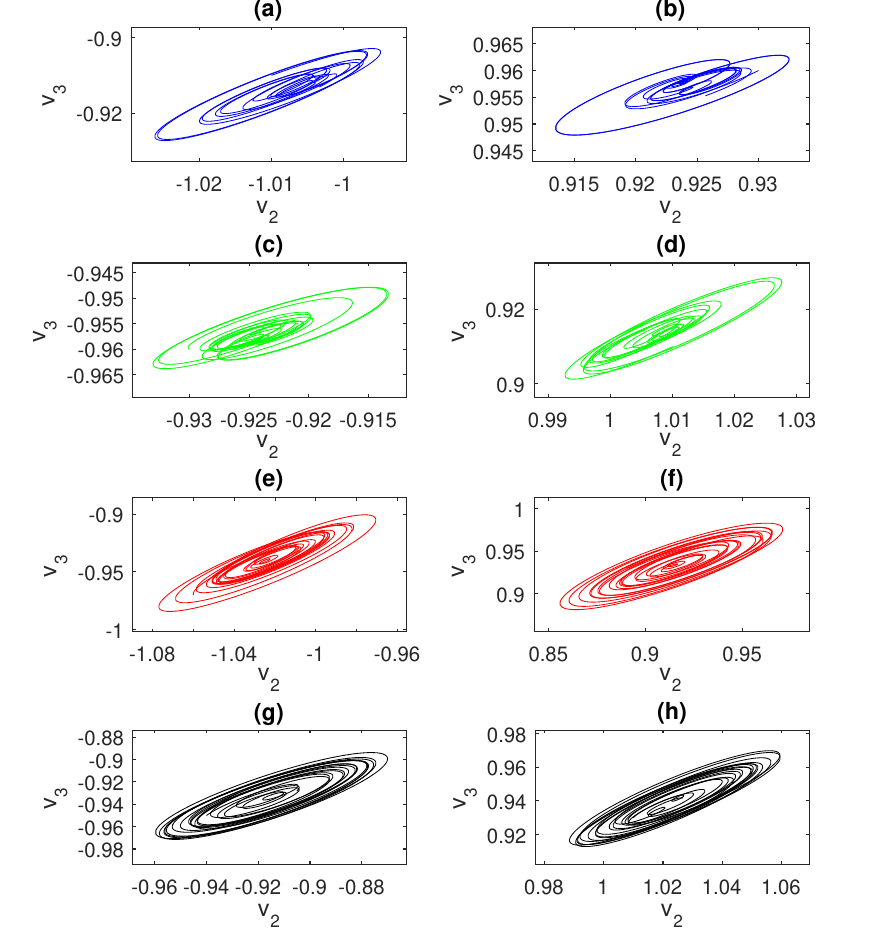}
\caption{Coexistence of eight chaotic attractors in the $12$-cell HNN (\ref{exdrive1})-(\ref{exresponse2})-(\ref{exresponse3})-(\ref{exresponse4}). The trajectories shown in (a)-(h) correspond to the initial data given in Table \ref{table2}. The figure displays that the doubling procedure is practical for chains of coupled HNNs.}
\label{fig7n}
\end{figure}

Next, let us focus on doubling the chaotic attractors of the network (\ref{exdrive1})-(\ref{exresponse2})-(\ref{exresponse3}) via another coupling. For that purpose, making use of the output of (\ref{exresponse3}) as an input for (\ref{exresponse1}) with $k=0.51$, we set up the response HNN
\begin{eqnarray} \label{exresponse4}
 \dot{v}_{1} &=& -v_{1}-1.4 \tanh(v_{1})+1.2 \tanh(v_{2})-7 \tanh(v_{3})+0.75 \tanh(z_{1}) + 0.013\tanh(z_{2}) \nonumber \\
 \dot{v}_{2} &=& -v_{2}+1.1 \tanh(v_{1})+2.8 \tanh(v_{3}) +0.007\tanh(z_1)+0.094\tanh(z_2) \nonumber \\ &+& 0.012\tanh(z_3) \nonumber \\
 \dot{v}_{3} &=& -v_{3}+0.51 \tanh(v_{1})-2 \tanh(v_{2})+4 \tanh(v_{3}) + 0.017 \tanh(z_1)  +0.008\tanh(z_2) \nonumber \\ &+& 0.089\tanh(z_3). 
\end{eqnarray}
In (\ref{exresponse4}), $z=(z_1,z_2,z_3)^T$ is an output of (\ref{exresponse3}). This time, (\ref{exresponse3}) is the drive whereas (\ref{exresponse4}) is the response network. For the aforementioned value of the parameter $k$ the attracting equilibrium points of (\ref{exresponse1}) are $(\pm 2.8555188, \mp 0.9640138, \mp 0.9388970).$ The skew product structure of the $12$-cell network (\ref{exdrive1})-(\ref{exresponse2})-(\ref{exresponse3})-(\ref{exresponse4}) leads to the formation of eight coexisting chaotic attractors in its dynamics. Figure \ref{fig7n} demonstrates the projections of each of them on the $v_2-v_3$ plane. Here, we utilize the same initial data shown in Table \ref{table1} for the first nine coordinates in case that the color of a trajectory coincides with the one represented in Figure \ref{fig6n}. The initial data for the remaining $v_1$, $v_2$ and $v_3$ coordinates are provided in Table \ref{table2}. The simulations depicted in Figure \ref{fig7n} confirm the coexistence of eight chaotic attractors in the dynamics of (\ref{exdrive1})-(\ref{exresponse2})-(\ref{exresponse3})-(\ref{exresponse4}), and their basins of attraction do not coincide. This manifests the feasibility of the doubling procedure to ensembles of unidirectionally coupled HNNs.
 
\begin{table}[ht]
\caption{Initial data of the trajectories represented in Figure \ref{fig7n}.}  
\centering  
\begin{tabular}{c c c c} 
\hline
& $v_1(0)$ & $v_2(0)$ & $v_3(0)$  \\  \hline
Figure \ref{fig7n}, (a) \ \ & $3.48$ & $-1.01$ & $-0.91$ \\  
Figure \ref{fig7n}, (b) \ \ & $-2.23$ & $0.93$ & $0.96$ \\  
Figure \ref{fig7n}, (c) \ \ & $2.22$ & $-0.93$ & $-0.96$ \\  
Figure \ref{fig7n}, (d) \ \ & $-3.49$ & $1.01$ & $0.91$ \\ 
Figure \ref{fig7n}, (e) \ \ & $3.63$ & $-1.06$ & $-0.97$ \\  
Figure \ref{fig7n}, (f) \ \ & $-2.12$ & $0.89$ & $0.93$ \\  
Figure \ref{fig7n}, (g) \ \ & $2.21$ & $-0.94$ & $-0.95$ \\  
Figure \ref{fig7n}, (h) \ \ & $-3.52$ & $1.01$ & $0.93$ \\  \hline  
\end{tabular} 
\label{table2} 
\end{table}

\section{Conclusions} \label{Sec5}

Researches on chaotic dynamics of artificial neural networks are interesting due to the crucial role in simulating brain dynamics as well as various application opportunities. This study is devoted to chaos analysis in unidirectionally coupled $3$-cell HNNs with two different dynamical structures in the absence of driving. Firstly, likewise the drive network, the response is supposed to possess a chaotic attractor when the coupling is not established. The permanency of chaotic behavior in the response is shown, and this phenomenon is called persistence of chaos. A remarkable novelty is its occurrence regardless of synchronization. This achievement provides an initial step for a mathematical background in the permanentness of chaos in dynamics of unsynchronized neuronal ensembles. Doubling of chaotic attractors is another concept that is revealed in the present paper. It opens up an opportunity to set up HNNs with multiple chaotic attractors. We demonstrate that if the response network admits two stable equilibrium points in the absence of the driving, then a $6$-cell HNN with two chaotic attractors is attained. Four and eight coexisting chaotic attractors are also shown. 

The presence of a bounded invariant region is required to confirm the sensitivity feature in the response dynamics, and this is deduced by the technique of Lyapunov functions. Indeed, it is used to affirm uniform ultimate boundedness. The presence of infinitely many unstable periodic outputs in a bounded region can be considered as one of the ingredients of chaos \cite{Wiggins88}, and an advantage of verifying uniform ultimate boundedness is the recognition of such outputs generated by the response network. In the future the obtained results can be developed for fractional order and memristive HNNs \cite{BCao23,Lin23} as well as for the ones under electromagnetic radiation \cite{Wan22}.

\section*{Appendix}

The proof of Theorem \ref{thm_sensitive} is as follows.

\noindent \textbf{Proof of Theorem \ref{thm_sensitive}.} The chaotic outputs of (\ref{drive}) are uniformly bounded, and they constitute an equicontinuous family of functions on the interval $[0,\infty)$ since their derivatives are also uniformly bounded. Accordingly, there is a positive number $M_1$ such that $\displaystyle\sup_{t\in\mathbb R}\left\|x(t)\right\| \leq M_1$ for every chaotic output $x(t)$ of (\ref{drive}). Moreover, because of the boundedness and invariantness of $\Lambda_y$, there is a positive number $M_2$ such that for every output $y(t)$ of the response network (\ref{response}) in which $x(t)$ is a chaotic output of (\ref{drive}) we have   $\displaystyle\sup_{t\in\mathbb R} \left\|y(t)\right\| \leq M_2$ whenever $y(0) \in \Lambda_y$.

Let a number $\widetilde{\delta}>0$, a point $\alpha_1 \in \Lambda_y$, and a chaotic output $x(t)$ of the drive HNN (\ref{drive}) be given. Owing to the sensitivity of the network (\ref{drive}), there exist numbers $\epsilon>0$, $\Delta >0$ and an interval $J\subset [0,\infty)$ with a length no less than $\Delta$ such that the inequality $\left\|x(t)-\widetilde{x}(t)\right\|>\epsilon$, $t\in J$, is satisfied for some chaotic output $\widetilde{x}(t)$ of (\ref{drive}) even though $x(0)$ and $\widetilde{x}(0)$ are arbitrarily close. It is worth noting that $\epsilon$ and $\Delta$ are both independent of $x(t)$ and $\widetilde{x}(t)$.
Fix a point $\alpha_2 \in \Lambda_y$ with $\left\|\alpha_1-\alpha_2\right\|<\widetilde{\delta}$. For the sake of brevity, in the rest of the proof we will denote $y(t)=\varphi_{x}(t,0,\alpha_1)$ and $\widetilde{y}(t)=\varphi_{\widetilde{x}}(t,0,\alpha_2)$. We will show the existence of positive numbers $\widetilde{\epsilon}$, $\widetilde{\Delta}$ and an interval $\widetilde{J}$ whose length is $\widetilde{\Delta}$ such that $\left\|y(t)-\widetilde{y}(t)\right\| \geq \widetilde{\epsilon}$ for every $t\in \widetilde{J}$.

The function $G:\mathbb R^3 \times \mathbb R^3 \to \mathbb R^3$ defined by $G(u,v) = g(u)-g(v)$ is uniformly continuous on the compact region of $\mathbb R^3 \times \mathbb R^3$ consisting of points $(u,v)$ with $\left\|u\right\| \leq M_1$ and $\left\|v\right\| \leq M_1$. Therefore, the collection of functions
$$
\mathcal{F} = \left\{g_j(x(t)) - g_j(\widetilde{x}(t)): \ j=1,2,3 \ \textrm{and} \ x(t), \widetilde{x}(t) \ \textrm{are chaotic outputs of HNN (\ref{drive})}  \right\}
$$
is an equicontinuous family on $[0,\infty)$. Accordingly, there exists a number $\tau >0$, which is less than $\Delta$ and does not depend on $x(t)$ and $\widetilde{x}(t)$, such that if $t_1$ and $t_2$ are nonnegative numbers satisfying $\left|t_1-t_2\right|<\tau$, then
\begin{eqnarray} \label{appendix1}
\left| \left( g_j\left(x(t_1)\right)-g_j\left(\widetilde{x}(t_1)\right) \right)  - \left( g_j\left(x(t_2)\right)-g_j\left(\widetilde{x}(t_2)\right) \right) \right| < \displaystyle \frac{L \epsilon}{2 \sqrt{3}}
\end{eqnarray}
for each $j=1,2,3$. 

Let us denote by $\eta$ the midpoint of the interval $J$. Thanks to the assumption $(A)$, the inequality
\begin{eqnarray*} \label{appendix2}
\left|g_k(x(\eta))-g_k(\widetilde{x}(\eta))\right| \geq \displaystyle \frac{L}{\sqrt{3}} \left\|x(\eta)- \widetilde{x}(\eta)\right\|>\displaystyle \frac{L \epsilon}{\sqrt{3}}
\end{eqnarray*}
is fulfilled for some integer $k$ with $1 \leq k \leq 3$. One can attain by means of (\ref{appendix1}) that
\begin{eqnarray*}\label{appendix3}
\left|g_k(x(\eta))-g_k(\widetilde{x}(\eta))\right|  - \left|g_k(x(t))-g_k(\widetilde{x}(t))\right|<\displaystyle \frac{L \epsilon}{2 \sqrt{3}}
\end{eqnarray*}
for every $t \in \left[\eta - \tau/2, \eta + \tau/2\right]$. Thus, for all such values of $t$ we have
\begin{eqnarray}\label{appendix4}
\left|g_k(x(t))-g_k(\widetilde{x}(t))\right| > \left|g_k(x(\eta))-g_k(\widetilde{x}(\eta))\right| - \displaystyle \frac{L \epsilon}{2 \sqrt{3}} > \displaystyle \frac{L \epsilon}{2 \sqrt{3}}.
\end{eqnarray}
It can be verified that
$$
\displaystyle \int_{\eta-\tau/2}^{\eta+\tau/2}\left(g(x(s))-g(\widetilde{x}(s))\right) ds
=\tau\left(  \left[ g_1(x(\theta_1))-g_1(\widetilde{x}(\theta_1))\right],   \left[ g_2(x(\theta_2))-g_2(\widetilde{x}(\theta_2))\right],   \left[ g_3(x(\theta_3))-g_3(\widetilde{x}(\theta_3))\right]\right)^T
$$
for some numbers $\theta_1$, $\theta_2$, and $\theta_3$ that belong to the interval $[\eta-\tau/2,\eta+\tau/2]$.
The inequality (\ref{appendix4}) implies that
\begin{eqnarray} \label{appendix5}
\Big\|\displaystyle \int_{\eta-\tau/2}^{\eta+\tau/2}\left(g(x(s))-g(\widetilde{x}(s))\right) ds \Big\| \geq \tau \left| g_k(x(\theta_k))-g_k(\widetilde{x}(\theta_k)) \right| > \displaystyle \frac{L \tau \epsilon}{2 \sqrt{3}}.
\end{eqnarray}

One can confirm for $\eta-\tau/2 \leq t \leq \eta + \tau/2$ that
\begin{eqnarray*}
 \left\|y\left( \eta + \frac{\tau}{2}\right)-\widetilde{y} \left(\eta + \frac{\tau}{2} \right)\right\|  
& \geq & \left|\lambda \right| \Big\| \displaystyle\int_{\eta - \tau/2}^{\eta + \tau/2} \left( g(x(s)) - g\left(\widetilde{x}(s)\right) \right) ds \Big\|
 - \left\|y\left( \eta - \frac{\tau}{2}\right)-\widetilde{y} \left(\eta - \frac{\tau}{2} \right)\right\| \\
 && - \displaystyle\int_{\eta - \tau/2}^{\eta + \tau/2} \big\|\widetilde{C}\big\| \left\|y(s) - \widetilde{y}(s)\right\| ds
 - \displaystyle\int_{\eta - \tau/2}^{\eta + \tau/2} \big\|\widetilde{W}\big\| \left\| f(y(s)) - f(\widetilde{y}(s))\right\| ds.
\end{eqnarray*}
Utilizing (\ref{appendix5}) we obtain the inequality
$$
\displaystyle \max_{t \in \left[\eta-\tau/2,\eta+\tau/2 \right]} \left\|y(t)-\widetilde{y}(t)\right\| \geq \displaystyle \frac{L \left|\lambda\right| \tau \epsilon}{2\sqrt{3} \big[ 2+ \big(c_0 +\big\|\widetilde{W}\big\| \big)\tau\big]},
$$
where $c_0=\max\left\{\widetilde c_1, \widetilde c_2, \widetilde c_3\right\}$.

Suppose that $\left\|y(t)-\widetilde{y}(t)\right\|$ attains its maximum on the interval $\left[ \eta-\tau/2, \eta + \tau/2 \right]$ at a point $\theta$ which belongs to that interval.
Let us define
$$ 
\widetilde{\epsilon} = \displaystyle \frac{L \left|\lambda\right| \tau \epsilon}{4 \sqrt{3} \big[2+\big(c_0+\big\|\widetilde{W}\big\|\big)\tau\big]}
$$
and
$$\widetilde{\Delta} = \displaystyle\min\bigg\{\displaystyle\frac{\tau}{2}, \frac{L \left|\lambda\right| \tau \epsilon}{8\sqrt{3} \big[2+\big(c_0+\big\|\widetilde{W}\big\|\big)\tau\big] \big[\big(c_0+\big\|\widetilde{W}\big\|\big)M_2 + \left|\lambda\right|M_3\big]} \bigg\},$$
in which
$M_3=\displaystyle \sup_{\nu \in\mathbb R^3} \left\|g(\nu)\right\|$. We would like to point out that $\widetilde{\epsilon}$ and 
$\widetilde{\Delta}$ are independent of the outputs $y(t)$, $\widetilde{y}(t)$. Moreover, the interval $\widetilde{J} = \big[ \zeta, \zeta+\widetilde{\Delta} \big]$ is a subset of $J$, where $\zeta = \theta$ if $\eta - \tau/2 \leq \theta < \eta$ and $\zeta = \theta - \widetilde{\Delta}$ if $\eta \leq \theta \leq \eta + \tau/2$. For $t \in \widetilde{J}$, the inequality
\begin{eqnarray*}
\left\|y(t) - \widetilde{y}(t)\right\| & \geq & \left\|y(\theta) - \widetilde{y}(\theta)\right\| - \Big| \displaystyle \int_{\theta}^t \big\| \widetilde{C}\big\| \left\|y(s) - \widetilde{y}(s)\right\| ds \Big| \\
&& - \Big| \displaystyle \int_{\theta}^t \big\| \widetilde{W}\big\| \left\|f(y(s)) - f(\widetilde{y}(s))\right\| ds \Big|
- \left|\lambda\right| \Big| \displaystyle \int_{\theta}^t  \left\|g(x(s)) - g(\widetilde{x}(s))\right\| ds \Big| \\
& \geq & \displaystyle \frac{L \left|\lambda\right| \tau \epsilon}{2 \sqrt{3} \big[2+\big(c_0+\big\|\widetilde{W}\big\|\big)\tau\big]}
-2\widetilde{\Delta} \left( c_0 M_2 +  \big\|\widetilde{W}\big\|M_2 +  \left|\lambda\right|M_3\right) \\ & \geq & \widetilde{\epsilon}
\end{eqnarray*}
is fulfilled. Consequently, the response HNN (\ref{response}) admits sensitivity. $\square$

\end{document}